\documentclass[final,lettersize]{siamltex}

\def\x{\mathbf{x}}

\def\y{\mathbf{y}}
\def\b{\mathbf{b}}
\def\c{\mathbf{c}}

\def\v{\mathbf{v}}

\def\z{\mathbf{z}}
\def\0{\mathbf{0}}
\def\1{\mathbf{1
}}

\def\A{\mathbf{A}}
\def\Y{\mathbf{Y}}

\def\lam{\lambda}

\def\r{{{\cal{R}}}}

\title{Benchmark Problems for Totally Unimodular Set System Auction}
\author{Ilan Adler
\thanks{Department of Industrial Engineering and Operations Research,
University of California, Berkeley, CA 94720
(adler@ieor.berkeley.edu).
}
\and Dorit S. Hochbaum
\thanks{Department of Industrial Engineering and Operations Research,
University of California, Berkeley, CA 94720
(hochbaum@ieor.berkeley.edu). This author's research was supported
in part by NSF award No. DMI-0620677 and CBET-0736232.}
}

\begin{document}
\maketitle

\pagestyle{myheadings} \markright{{\sl }}

\begin{abstract}
We consider  a generalization of the $k$-flow set system auction where the set to be procured by a
customer corresponds to a feasible solution to a linear programming problem where the
coefficient matrix and right-hand-side together constitute a totally unimodular matrix.  Our results generalize and strengthen bounds
identified for several benchmarks, which form a crucial component in the study of
frugality ratios of truthful auction mechanisms.
\end{abstract}



\pagestyle{myheadings} \thispagestyle{plain} \markboth{I. Adler and D. S. Hochbaum
}{Benchmark Problems for Totally Unimodular Set System Auction}

\section{Introduction} \label{section:intro}
In a {\it set system auction}, a customer procure the services of a subset of a set of available agents, each of whom quotes a specific amount (called a {\it bid}) for the service.   The customer must select from predetermined {\it feasible subsets} of the agents, while minimizing total payments to the agents.

Several authors (\cite{kkt,cegp,ck,gc,ksm}) have considered variants of set system auction problems in which the feasible sets are defined in the context of networks where nodes or edges represent agents and the feasible sets are characterized by network structures such as spanning trees;
single and  multi-paths; vertex covers and cuts .
An {\it auction mechanism} determines a selection scheme -- that is, how to decide on the set of
winning agents, and how much to pay each of the winning agents,
although the mechanism can devise prices different from the bid values of the agents.  A desirable property of the mechanism is that is should be designed to be {\em truthful} so that agents have no incentive to lie about their true cost of providing service while quoting their bids.

This truthfulness feature results in premium payments to
the agents that exceeds the agents' truthful bids, similar to the single agent case where the agent with the winning bid can raise her bid up to the second lowest bid without affecting her position.  A key goal in designing auction mechanism is to minimize the additional costs to the customer resulting from seeking truthful bids. A
major theme within this stream of research focuses on the notion of frugality that was introduced in \cite{at} and extended by \cite{Tal}. The {\it frugality ratio} measures the
ratio of the payments made by a truthful auction mechanism as compared to some benchmark which is intended to represent the least cost auction mechanism that assures truthful biding. In \cite{e2g}, two frequently discussed benchmarks related to bids satisfying Nash equilibrium within the context of several set auctions related to networks are analyzed. Both problems are defined as optimization problems which share the same set of constraints and objective. In the first benchmark problem, which was introduced in \cite{ksm}, the objective function is minimized, while in the second problem, which was introduced in \cite{e2g}, the objective function is maximized.

Two recent papers, \cite{cegp} and \cite{ksm}, present and discuss set auctions in which the
feasible subset corresponds to $k$ disjoint $s-t$ paths, which can be found by solving a min cost $k$-flow problem (and thus the corresponding auction problem is commonly called the  {\it $k$-flow auction problem}). In particular, they prove several results related to the two benchmark problems (associated with the $k$-flow auction problem) as discussed above.
Our goal in this paper is to present a generalization of the  $k$-flow  auction problem to a linear programming (LP) problem whose coefficient matrix and right-hand-side together constitute a totally unimodular matrix. In particular, in this more general setting, we provide novel simple proofs based on standard LP theory for (and extend in one case) all the bounds related to the two associated benchmark problems. 

\subsection{Contributions}
Specifically, we propose a set system auction framework that generalizes the $k$-flow auction problem.  The feasible sets are defined in terms of a partition of the columns of a totally unimodular matrix.  

We denote by $\mu(k)$ ($\nu(k)$) the optimal objective function value of the maximization (minimization) benchmark problem. The formal definition of these benchmark problems is
given in Section \ref{sec:bench}. In \cite{cegp}, it is proved that for the $k$-flow auction problem $\mu(k) \geq k [\phi(k+1) -
\phi(k)]$. We prove that this inequality is in fact an {\em equality} for the $k$-flow
auction problem, and the equality applies in the more general setting of total unimodularity.

Both \cite{ksm} and \cite{cegp} present a pruned $k$-flow auction problem which plays a key rule in analyzing the associated frugality ratios. Specifically, the {\it pruned $k$-flow auction problem}
is defined as a restriction on the $k$-flow auction problem in which all edges who are not in the min cost solution of the original $k$-flow auction problem are removed. We denote by  $\tilde{\mu}(k)$  the optimal objective function value of the maximization benchmark problem associated with the pruned problem.  The precise
definition of the pruned problem is given in Section \ref{sec:max}. For the $k$-flow auction problem, it is proved in \cite{ksm} that $\mu(k) \geq
\frac{1}{k+1}\tilde{\mu}(k)$, while it is proved in \cite{cegp} that $\mu(k) \geq
\tilde{\mu}(k)$.  We prove, for the more general case of total unimodularity, the same
lower bound for $\mu(k)$ as in \cite{cegp}. We also provide an upper bound for $\mu(k)$
in terms of $\tilde{\mu}(k)$.

\cite{cegp} presents a lower bound for $\nu(k)$. The key to this bound is a theorem, first introduced in \cite{gc},
related to a minimum cost $k$-flow problem with the property that removing any single edge from the
underlying network does not affect the total minimum cost. We present a simpler proof of this
theorem generalized to totally unimodular matrices. We then proceed to use this theorem
to prove a lower bound for $\nu(k)$, which generalizes a similar bound established in
\cite{cegp} for the $k$-flow case.

\subsection{Notation and Preliminaries}
 Matrices are denoted by boldface, upper case fonts, e.g.\
$\A$. Vectors are denoted by boldface, lower case fonts, e.g.\ $\v$.

For an $n\times m$ matrix $\A$, let $J=\{ 1,\ldots ,n\}$ be the set of columns of the
matrix.  Let $J'\subseteq J$ then $\A _{J'}$ indicates the $|J'|\times m$ submatrix of
$\A$ restricted to the set of columns $J'$.  Similarly, $\x _{J'}$ indicates a
restriction of the vector $\x \in \r ^n$ to a $|J'|$-vector.

\section{First Price Set System Auction Problem}

In the {\it first price set system auction} each agent bids her true cost and the customer
selects a feasible set with minimum total cost. Specifically:
given an $m\times n$ matrix $\A$, an $m$ vector $\b$, a  nonnegative $n$ vector $\c$, and a positive
integer $k$, the goal is to find $k$ mutually exclusive subsets
 $J_1, \ldots, J_k$  of $\{1, \ldots, n\}$ whose total cost is minimized. That is:
\begin{eqnarray}
\label{pr:set_auction}
\min \;\; \sum_{i =1}^{k} \sum_{ j \in J_i} \c_j \hspace{24mm} \\
\label{eq:set_auction}
 \mbox{ subject to }\;\;\sum_{ j \in J_i} \A_j  = \b,\;\; i=1, \ldots, k . \nonumber
 \end{eqnarray}
\medskip

 We assume throughout the paper that
  the matrix $(\A, \b)$ is totally unimodular and that $\b \neq \0$. We show below that under these assumptions, it is possible to present problem (\ref{pr:set_auction}) as the following
  linear program:
  \[  \hspace{.4in}\begin{array}{lll}
 {\cal{P}}(k):\;\; \;\phi(k) = & \hspace{10mm} \min  & \c \x\\
 &\mbox{subject to } \overline{} &  \A \x =k \b, \;\; \0 \leq \x \leq \1,\\
\end{array}
\]
where $k$ is a positive integer, $\c$ is a row vector, $\b,\x$ are column vectors,
$(\A, \b)$ is totally unimodular, and $\b \neq \0$.

\medskip

The $k$-flow set auction problem, which is discussed in detail in \cite{ksm} and
\cite{cegp}, is a special case of ${\cal{P}}(k)$. Specifically, given a directed graph
$G$ with an origin node $s$, a sink node $t$, and a capacity $1$ for each edge, we can
set $\A$ as the node-edge incidence matrix of $G$ and $\b$ as a vector of zeros with the
exception of $1$ in the row corresponding to $s$ and $-1$ in the row corresponding to
$t$.  The resulting matrix, $(\A, \b)$, is totally unimodular.
\medskip

The total unimodularity of $\A$ guarantees that an integer (and thus a binary) optimal
solution to ${\cal{P}}(k)$ exists whenever $\b$ is an integer vector and the problem is
feasible. The decomposition of this solution to $k$ binary feasible solutions to
${\cal{P}}(1)$, as required by (\ref{pr:set_auction}), is possible due to the following
proposition, which is a variant of the {\it integral decomposition property} theorem of
Baum and Trotter \cite{bt}.
\medskip
\begin{proposition}
\label{pr:trotter}
 Let $\bar{\x}$ be a binary feasible solution to ${\cal{P}}(k)$. Then,
 there exist $k$ binary feasible solutions ${\bar{\x}}^1, \ldots, {\bar{\x}}^k$ to
 ${\cal{P}}(1)$ such that
 ${\bar{\x}} = \sum_{i=1}^k {\bar{\x}}^i$.
\end{proposition}
\medskip
\begin{proof}
  The proof follows by induction. The case $k=1$ is trivial. Suppose the proposition is true for $k-1 \geq 2$.
  Let $J = \{ j \;| \;{\bar{\x}}_j =1 \}$. Since $\frac{1}{k}\bar{\x}$ is a feasible solution
  to ${\cal{P}}(1)$ and since $\A$ is totally unimodular, there exists a binary vector
  $\x^k$ such that $\A_J \x^k_J = \b$. Let $\hat{\x} = \bar{\x} - \x^k$. Then,
  \[
  \A \hat{\x} = \A (\bar{\x} - \x^k)= k\b -\b= (k-1)\b, \;\;
  \0 \leq \hat{\x} \leq \1, \;\; \hat{\x} \mbox{ is binary.}
  \]

By the induction assumption, there exist $k-1$ binary vectors ${\hat{\x}}^i, \; i=1,
\ldots, k-1$ such that
 $\hat{\x} = \sum_{i=1}^k {\hat{\x}}^i$ and for each $i$,
 ${\hat{\x}}^i$ is a feasible solution to ${\cal{P}}(1)$. Letting
 ${\bar{\x}}^i = {\hat{\x}}^i$ for $i=1, \ldots, k-1$, and ${\bar{\x}}^k =\x^k$  completes the proof.
 \qquad
\end{proof}

\medskip

It will be useful to consider ${\cal{P}}(\lam)$ for real nonnegative parameter value $\lam$. In this case, we can present
the problem as a parametric linear program as follows:
\begin{equation}
\label{eq:para_lp}
   \hspace{.4in}\begin{array}{lll}
 {\cal{P}}(\lam):\;\; \;\phi(\lam) = & \hspace{10mm} \min  & \c \x\\
 &\mbox{subject to } \overline{} &  \A \x =\0 +\lam\b, \;\; \0 \leq \x \leq \1.\\
\end{array}
\end{equation}


It is well known that the function $\phi (\lam)$  is a continuous piecewise linear
convex. In addition, if the problem has an optimal solution for at least one parameter
value $\lam$, there exist {\it break points} $\lam_0, \lam_1, \ldots, \lam_{\ell}$ and $
h_i , g_i$ real numbers $i=1,\ldots ,\ell$,  such that,
 \begin{eqnarray}
 \label{eq:para_1}
 {\cal{P}}(\lam)\;
 \mbox{ is feasible for } \lam \in [\lam_0, \lam_{\ell}], \hspace{27mm}\\
 \label{eq:para_2}
 \phi(\lam) = h_i + g_i \lambda \; \mbox{ for }
  \lambda_{i-1} \leq \lambda \leq \lambda_{i}, \;\;
  i=1, \ldots, \ell. \hspace{3mm}
   \end{eqnarray}
And, since the matrix $(\A, \b)$ is totally unimodular,
  \begin{eqnarray}
 \label{eq:para_3}
  \mbox{all the break points } \lam_i \mbox{ are integers}.
  \hspace{20mm}
    \end{eqnarray}

 We assume, by introducing lexicographical ordering of the variables if necessary, that  ${\cal{P}}(k)$ has a unique optimal solution $x^*$.
 We also assume that ${\cal{P}}(k)$ is {\it monopoly-free}. That is, for every $j=1, \ldots, n$,
 there exists a feasible solution $\x$ to ${\cal{P}}(k)$ with $\x_j=0$. This property, as demonstrated in the next
 proposition, implies that there exists a feasible solution to  ${\cal{P}}(k+1)$.

 \medskip

 \begin{proposition}
\label{pr:p_K_plus_1}
 Suppose that for every $j=1, \ldots, n$, ${\cal{P}}(k)$ has a feasible solution
 ${\bar{\x}}^j$ with ${\bar{\x}}^j_j =0$.
  Then, there exists a feasible solution to ${\cal{P}}(k+1)$.
\end{proposition}
\medskip
\begin{proof}
Let $\bar{\x} = \frac{1}{n} \sum_{j=1}^n {\bar{\x}}^j$. Then, since $\bar{\x}$ is a
strict convex combination of feasible solutions to ${\cal{P}}(k)$, and since
${\bar{\x}}^j_j =0$ for every $j=1, \ldots, n$, we have that $\bar{\x}$ is a feasible
solution to ${\cal{P}}(k)$ and that $\0 \leq \bar{\x} < \1$. Hence, there exists a
sufficiently small positive $\bar{\epsilon}$ such that  for $ \lam \in [k,
(1+\bar{\epsilon})k]$, $\left(\frac{\lam}{k}\right)\bar{\x}$ is a feasible solution to
${\cal{P}}(\lam)$. Recalling that by (\ref{eq:para_3}), all the breakpoints $\lam_i$ of $\phi(\lam)$ in the parametric LP (\ref{eq:para_lp}) are
integers,  and considering (\ref{eq:para_1}),
we have that ${\cal{P}}(k+1)$ is feasible. \qquad
\end{proof}

  \medskip

Throughout the rest of the paper we make frequent use of the {\it optimality conditions}
of ${\cal{P}}(\lambda)$ that are stated in the following proposition.

   \medskip

  \begin{proposition}
  \label{pr:opt_cond}
  Let $\bar{\x} = ({\bar{\x}}_{J_0}, {\bar{\x}}_{J_f}, {\bar{\x}}_{J_1})$,  where
  ${\bar{\x}}_{J_0} = \0, \; \0 < {\bar{\x}}_{J_f} < \1$, and ${\bar{\x}}_{J_1} = \1$,
  be a feasible solution to ${\cal{P}}(\lam)$. Then, $\bar{\x}$ is optimal  if, and only if, there exists
  an m-row vector $\bar{\y}$ such that,
  \[
   \bar{\y} \A_{J_0} \leq \c_{J_0},\;\; \bar{\y} \A_{J_f} = \c_{J_f},\;\;
    \bar{\y} \A_{J_1} \geq \c_{J_1}.
  \]
  In addition, $ \phi(\lam) = \c {\bar{\x}} =
   \lam \bar{\y} \b +  (\c_{J_1} - \bar{\y} \A_{J_1}) \1$.
 \end{proposition}

 \medskip

  \begin{proof}
   The proof follows directly by setting the dual problem to ${\cal{P}}(k)$ and imposing
   the complementary slackness conditions. \qquad
  \end{proof}

\section{Benchmark Problems} \label{sec:bench}

 We associate two benchmark problems with ${\cal{P}}(k)$.
 Let $\x^*$ be the unique optimal solution to ${\cal{P}}(k)$, and let
 \[
 J_0 = \{ j\; | \; \x^*_j =0\},\;\;
 J_1 = \{ j\; | \; \x^*_j =1\}.
 \]
 In both benchmark problems, we consider changing the prices
 $\c_j$ to $\z_j$
  while satisfying the following requirements
 (where the decision variables form a row $n$-vector $\z$):
  \begin{eqnarray}
  \label{eq:bench_1}
 \z_{J_1} \geq \c_{J_1}, \;\;
 \z_{J_0} = \c_{J_0}, \hspace{78mm}\\
 \label{eq:bench_2}
  \z  \x^* \leq \z \bar{\x}  \hspace{1mm} \mbox{ for every binary feasible solution } \bar{\x}
  \mbox{ to } {\cal{P}}(k), \hspace{28mm}\\
  \label{eq:bench_3}
  \mbox{For every } j=1, \ldots, n,
   \mbox{ there exists a binary feasible solution } {\bar{\x}}^j
    \mbox{ to } {\cal{P}}(k) \\
    \mbox{such that } {\bar{\x}}^j_j=0 \mbox{ and } \z\x^* = \z \bar{\x}^j.
    \hspace{60mm}
    \nonumber.
 \end{eqnarray}	


\medskip

The two benchmark problems share the same objective function
 $\z_{J_1} \x^*_{J_1}$, but differ by whether to maximize or minimize it
 (note that since $\x^*_{J_1}=\1$, we can refer to the objective function as
 $\z_{J_1} \1$). In the {\it max benchmark problem}, ${\cal{B}}_{max}(k)$
that was first introduced in \cite{e2g}, the objective is to maximize the objective
function, while the objective in the {\it min benchmark problem}, ${\cal{B}}_{min}(k)$
that was first introduced in \cite{kkt}, is to minimize the objective function. We denote
the optimal objective value of the max benchmark problem by $\mu(k)$, and the optimal
objective value of the min benchmark problem by $\nu(k)$.

 \medskip
\subsection{The Max Benchmark Problem} \label{sec:max}

As was pointed out in \cite{e2g}, and can easily be verified, any optimal solution to the
max benchmark problem satisfies the requirement stated in (\ref{eq:bench_3}). Therefore,
if a separation oracle for condition (\ref{eq:bench_2}) is provided, the max benchmark
problem can be formulated as a linear program. In \cite{ksm} and \cite{cegp}, the max
benchmark problem for the special case of $k$-flow is analyzed by using elaborate network
theory results. In contrast, by considering Proposition \ref{pr:opt_cond}, we present an
explicit LP formulation of the max benchmark problem for the more general case as
presented in (\ref{pr:set_auction}). We then proceed, by using standard LP theory, to
provide simpler proofs that generalize the results in \cite{ksm} and \cite{cegp}.

 \medskip

Applying directly the optimality conditions of Proposition \ref{pr:opt_cond}, and
considering requirements (\ref{eq:bench_1}) and (\ref{eq:bench_2}), the max benchmark
problem can be formulated as the following linear program:

  \medskip

\[
 \hspace{.2in}\begin{array}{lll}
 {\cal{B}}_{max}(k):\;\; \; \mu(k) = & \hspace{9mm} \max  & \z_{J_1} \1\\
 &\mbox{subject to } \overline{} & \y  \A_{J_0}  \leq  \c_{J_0} \\
 & &  \y  \A_{J_1}  \geq \z_{J_1} \\
  & &  \hspace{5mm} \z_{J_1}   \geq  \c_{J_1} \\
\end{array}
\]

In \cite{cegp}, it is proved that for the $k$-flow problem, $\mu(k) \geq k [\phi(k+1) -
\phi(k)]$. In the following theorem we prove that this inequality is in fact an equality
even in the more general case of total unimodularity.
   \medskip

\begin{theorem}
\label{th:para_1}
 $\mu(k) = k [\phi(k+1) - \phi(k)]$.
\end{theorem}
\\
\begin{proof}
 Observing that at optimality, $\y  \A_{J_1}  = \z_{J_1}$, and that
 $\A_{J_1} \1 = k\b$, ${\cal{B}}_{max}(k)$ can be written as:
\[
 \mu(k)=\max_{\y \in \Y(k)} k \y \b = k \beta(k),
\]
where
\[
 \Y(k) = \{ \y \; | \; \y  \A_{J_0}  \leq  \c_{J_0}, \;\; \y  \A_{J_1}  \geq  \c_{J_1}\},
 \; \mbox{ and } \; \beta(k) = \max_{\y \in \Y(k)} \y \b.
\]


Considering ${\cal{P}}(k)$ as a parametric LP as in (\ref{eq:para_lp}), noting
 that $\Y(k)$ is the set of all optimal solutions
 to the dual of ${\cal{P}}(k)$, and applying a well known result about parametric LP
 (see e.g. Theorem 8.2 of \cite{m}),
 we have that
 $\beta(k) = g_i$, where $\lam_{i-1} \leq k  < \lam_i$.  However, since
 by (\ref{eq:para_3})
 the breakpoints $\lam_i$ are all integers,  $g_i= \phi (k+1
)- \phi (k )$, which implies that
$\mu(k)=k \beta(k)=k[\phi (k+1)- \phi (k )].$ \qquad
  \end{proof}
\\

The {\it pruned first price set system auction} problem associated with ${\cal{P}}(k)$ is
constructed by deleting all the columns which are not used in the optimal solution of  ${\cal{P}}(k+1)$.
Specifically, let $\bar{\x}$ be the optimal solution of ${\cal{P}}(k+1)$ (we assume it is unique), and let
$J = \{ j \;| \;{\bar{\x}}_j =1 \}$. Then,
 the   pruned first price set system auction problem is defined as:
\[
 \hspace{.4in}\begin{array}{lll}
 {\tilde{\cal{P}}}(k):\;\; \; \tilde{\phi}(k) = & \min  & \tilde{\c} \tilde{\x}\\
 &\mbox{subject to } \overline{} &  \tilde{\A} \tilde{\x} =k \b, \;\; \0 \leq \tilde{\x} \leq \1\\
\end{array}
\]
where $\tilde{\c} = \c_{J}$, $\tilde{\A} = \A_{J}$, and $\tilde{\x}$ is a  $|J_1|$ column vector.

\medskip

We denote by $\tilde{\mu}(k)$ the optimal objective function of the max benchmark problem
associated with the pruned problem, ${\tilde{\cal{P}}}(k)$. Since the pattern of the
auction's agents' behavior is simpler for the pruned problem than it is for the original
problem, it is hoped that the values of the associated $\mu(k)$ and $\tilde{\mu}(k)$ are
close.

Indeed, for the $k$-flow problem, it is proved in \cite{ksm} that $\mu(k) \geq
\frac{1}{k+1}\tilde{\mu}(k)$, while a stronger lower bound result is proved in
\cite{cegp} -- that $\mu(k) \geq \tilde{\mu}(k)$. In the following theorem, we prove for
the more general case of total unimodularity, the stronger lower bound for $\mu(k)$ (as
in \cite{cegp}). We also provide an upper bound for $\mu(k)$ in terms of
$\tilde{\mu}(k)$.

\medskip

\begin{theorem}
\label{th:main}
 $\tilde{\mu}(k) \leq \mu(k) \leq (k+1)\tilde{\mu}(k)$.
\end{theorem}\\
{\em Proof}.
\begin{description}
\item[(a)] \hspace{2mm} Note that $\phi(k+1) = \tilde{\phi}(k+1)$. Secondly, since
    ${\tilde{\cal{P}}}(k)$ is a restriction of ${\cal{P}}(k)$,  $\phi(k) \leq
    \tilde{\phi}(k)$. Thus, by Theorem \ref{th:main},
  \[
  \mu(k) = k[\phi (k+1)- \phi (k )] \geq   k[\phi(k+1)- \tilde{\phi}(k)] = \tilde{\mu}(k)
  \]
This completes the proof of the lower bound.
  \item[(b)] \hspace{2mm} Let $\bar{\x}$ be a
   binary optimal solution to ${\cal{P}}(k+1)$. By Proposition \ref{pr:trotter},
      there exist binary ${\bar{\x}}^i, i=1, \ldots, k+1$  such
      that
$ \A{\bar{\x}}^i = \b$ and $\bar{\x}=\sum_{i=1}^{k+1} {\bar{\x}}^i$. Let $\delta =
\max_{1 \leq i \leq  k+1} \c \bar{\x}^i$. Then,
  \[
   \tilde{\mu}(k) =  k[\phi (k+1)- \tilde{\phi}(k)] \geq
   k[\phi (k+1)- (\phi (k+1) -\delta)] =
    k \delta.
     \]
  and
  \[
    k \delta \geq k \frac{\phi (k+1)}{k+1}
    \geq  \frac{k[\phi (k+1) - \phi (k)]}{k+1} = \frac{\mu(k)}{k+1}.
    \qquad \qquad \endproof
  \]
 \end{description}

\subsection{The Min Benchmark Problem}

Let $\nu(k)$ be the optimal value of the min benchmark problem. Unlike the max benchmark
problem, the third requirement of the min benchmark problem (\ref{eq:bench_3}), is not
redundant.  Moreover, it is shown in \cite{ck} that the min benchmark problem for the
special case of $k$-flow is NP-complete even for $k=1$. However, paper \cite{cegp}
presents a lower bound for $\nu(k)$ (for the $k$-flow case). The key to this bound is a
theorem, first introduced in \cite{gc}, which is related to a minimum cost $k$-flow
problem with the property that removing any single edge from the
underlying network does not affect the total minimum cost. In the following proposition, we present a simpler proof of the same theorem but in our more general setup.

\medskip

\begin{proposition}
\label{pr:B1}
 Suppose that for every $j=1, \ldots, n$, ${\cal{P}}(k)$ has a binary optimal solution
 ${\hat{\x}}^j$ with ${\hat{\x}}^j_j =0$.
  Then, there exist $k+1$ binary optimal solutions ${\bar{x}}^1, \ldots, {\bar{x}}^{k+1}$
  to ${\cal{P}}(1)$ such that
   $\sum_{i=1}^{k+1} {\bar{x}}^i$ is an optimal solution to ${\cal{P}}(k+1)$.
\end{proposition}

\medskip

\begin{proof}
Let $\hat{\x}$ be a strict convex combination of all the binary optimal solutions
${\hat{\x}}^j$, $j=1,\ldots ,n$, to ${\cal{P}}(k)$. Let
  \[
  J_f = \{ j \;| \; {\hat{\x}}_j >0 \},\;\; J_0 = \{ j \;| \; {\hat{\x}} =0 \}.
  \]
By the assumption of the proposition and the definition of $J_f$, we have that
  $ \0\ < {\hat{\x}}_{J_f} < \1$. Since ${\hat{\x}}$ is an optimal solution to ${\cal{P}}(k)$, and by Proposition
  \ref{pr:opt_cond}, there exists $\bar{\y}$ such that
\[
 \bar{\y} \A_{J_0} \leq \c_{J_0}, \;\;  \bar{\y} \A_{J_f} = \c_{J_f}.
 \]
  Moreover,  Proposition \ref{pr:opt_cond} implies that for nonnegative $\lam$,
  \begin{equation}
  \label{eq:opt_intr}
  \x_{J_0} =\0, \;\; \0 \leq \x_{J_f}  \leq \1, \;\; \A_{J_f} \x_{J_f} = \lam\b,\;
  \Rightarrow  \x\; \mbox{ is optimal for } {\cal{P}}(\lam).
  \end{equation}
  Now, consider ${\cal{P}}_f(\lam)$, the parametric LP (\ref{eq:para_lp})
   restricted to $J_f$. That is,
\begin{equation}
   \hspace{.4in}\begin{array}{lll}
 {\cal{P}}_f(\lam):\;\; \;\phi_f(\lam) = & \hspace{10mm} \min  & \c_f \x_f\\
 &\mbox{subject to } \overline{} &  \A_f \x_f =\0 +\lam\b, \;\; \0 \leq \x_f \leq \1.\\
\end{array}
\end{equation}
 Note that the proposition's assumption implies that ${\cal{P}}_f(k)$ is {\it monopoly-free}.
 That is, for every $j \in J_f$ there exists a binary feasible solution $\x$ to
 ${\cal{P}}_f(k)$ with $\x_j=0$. Thus,
  applying Proposition \ref{pr:p_K_plus_1} to ${\cal{P}}_f(k)$, we have
 that there exists a binary feasible solution $\bar{\x}_{J_f}$
 to ${\cal{P}}_f(k+1)$. Now, let
 $\bar{\x} = (\bar{\x}_{J_0}, \bar{\x}_{J_f}) $ where
 $ \bar{\x}_{J_0} = \0$. We define
  \[
  \bar{\x}(\lam) =
  \left(\frac{\lam}{k+1}\right)\bar{\x}.
 \]

 Since $\bar{\x}(\lam)$ is feasible for ${\cal{P}}(\lam)$ for
  $\lam \in  [0, k+1]$, and by (\ref{eq:opt_intr}), we have that
  $\bar{\x}(\lam)$ is optimal for ${\cal{P}}(\lam)$ for $\lam \in  [0, k+1]$. So,
  for $\lam \in  [0, k+1]$,
  \[\phi(\lam) =
  \c \bar{\x}(\lam) = \c_{J_f} \left(\frac{\lam}{k+1}\right)\bar{\x}_{J_f}
  = \left(\frac{\lam}{k+1}\right) \c_{J_f} \bar{\x}_{J_f} =
  \left(\frac{\lam}{k+1}\right)\phi(k+1).
  \]
  Thus, $\phi(1)=\frac{1}{k+1} \phi(k+1)$, which implies that
  $\phi(\lam)=\lam \phi(1)$.
  Finally, recalling that $\bar{\x} = \bar{\x}(k+1)$ is a binary
optimal solution to
   ${\cal{P}}(k+1)$, and
   by Proposition \ref{pr:trotter},
   there exist $k+1$ binary feasible solutions ${\bar{\x}}^1, \ldots, {\bar{\x}}^{k+1}$
  to ${\cal{P}}(1)$ such that
   $\bar{x} = \sum_{i=1}^{k+1} {\bar{x}}^i$.
    Thus,
   \[
   0=\phi(k+1) - (k + 1) \phi(1) = \c \bar{\x} - (k + 1) \phi(1) =
   \c ( \sum_{i=1}^{k+1} {\bar{\x}}^i) - (k + 1) \phi(1)
   = \sum_{i=1}^{k+1} (\c{\bar{\x}}^i - \phi(1))
   \]
   Noting that $\c{\bar{\x}}^i - \phi(1) \geq 0$ for $i=1, \ldots, k+1$, and
    $\sum_{i=1}^{k+1} (\c{\bar{\x}}^i - \phi(1))=0$,
    we conclude that for $i=1, \ldots, k+1$, $\c{\bar{\x}}^i =\phi(1)$. Hence, ${\bar{\x}}^i\; (i=1, \ldots, k+1)$ are optimal for  ${\cal{P}}(1)$.
   \qquad
 \end{proof}

\medskip

  Based on Proposition \ref{pr:B1}, we conclude this section by proving a lower bound for
  $\nu(k)$ which is a straightforward generalization of a similar bound established
  in \cite{cegp} for the $k$-flow case.

  \medskip

   We call a set $S=\{\x^1, \ldots, \x^{k+1}\}$ a {\it feasible collection} for ${\cal{P}}(k+1)$,
   if for
   \newline \noindent $i=1, \ldots, k+1$, $\x^i$ is a binary feasible solution to ${\cal{P}}(1)$
    and $\sum_{i=1}^{k+1} \x^i$ is a feasible solution to
   ${\cal{P}}(k+1)$.
    We define
    $\gamma_{\c}(S)= \max_{\x^i \in S} \c \x^i$ and
    $\Gamma_{\c}=\min_{S \in \Omega} \gamma_{\c}(S)$, where
    $\Omega$ denotes the set  of all feasible collections $S$ for
    ${\cal{P}}(k+1)$.\\

    Let $\bar{\z}$ be an optimal solution to the min benchmark problem. We denote
    by ${\cal{P}}_{\bar{\z}}(k)$ problem ${\cal{P}}(k)$ where $\bar{\z}$ replaces
    $\c$ in the objective function, and by $\phi_{\bar{\z}}(k)$ the optimal objective function value of ${\cal{P}}_{\bar{\z}}(k)$.

   \medskip

\begin{theorem}
\label{th:minbench}
 $\nu(k) \geq  k \Gamma_{\c} $.
\end{theorem}
\medskip

{\em Proof}.
 Let $x^*$ be the unique
 binary optimal  solution to ${\cal{P}}(k)$. Then, by requirement
 (\ref{eq:bench_2}) of the min benchmark problem, we have that
 $x^*$ is a binary optimal  solution to ${\cal{P}}_{\bar{\z}}(k)$.
 However, requirement
(\ref{eq:bench_3}) of the min benchmark problem satisfies the assumption of Proposition \ref{pr:B1},
 with respect to ${\cal{P}}_{\bar{\z}}(k)$. Thus, there exist $k+1$ binary optimal solutions ${\bar{x}}^1, \ldots,
{\bar{x}}^{k+1}$ to ${\cal{P}}_{\bar{\z}}(1)$ such that $\sum_{i=1}^{k+1} {\bar{x}}^i$ is
an optimal solution to ${\cal{P}}_{\bar{\z}}(k+1)$. It also follows, by the convexity of $\phi(\lam)$, that
$  k\phi_{\bar{\z}}(k) = k \phi_{\bar{\z}}(1)$.
Thus, noting that for $i=1, \ldots, k+1$, $\z \x^i = \phi_{\z}(1)$,
 and that (\ref{eq:bench_1}) implies that for all $S \in \Omega$, $\gamma_{\z}(S) \geq \gamma_{\c}(S)$,
 we have
\[
   \nu(k)=\bar{\z}\x^* = \phi_{\z}(k) = k\phi_{\bar{\z}}(1) =
   k \gamma_{\z}({\bar{x}}^1, \ldots, {\bar{x}}^{k+1})
    \geq k \gamma_{\c}({\bar{x}}^1, \ldots, {\bar{x}}^{k+1})
   \geq    k \Gamma_{\c}.
    \qquad \endproof
\]

 \end{document}